# Fault Tolerant Super Twisting Sliding Mode Control of a Quadrotor UAV Using Control Allocation

Mehmet Karahan [a,1,*], Mertcan Inal [a,2], Cosku Kasnakoglu [a,3]

[a] TOBB University of Economics and Technology, Electrical and Electronics Engineering Department, Sogutozu Street No: 43, Cankaya, Ankara 06510, Turkey
[1] m.karahan@etu.edu.tr; [2] mertcan.inal@etu.edu.tr; [3] kasnakoglu@etu.edu.tr
* Corresponding Author

ARTICLE INFO | ABSTRACT

In this study, a fault-tolerant super-twisting sliding mode controller with a control allocation system for a quadrotor aircraft is proposed. Super twisting sliding mode control is a robust control technique that handles a system with a relative degree equal to one. A super-twisting sliding mode controller is proposed because of its robustness to uncertainties and perturbations. It increases accuracy and reduces chattering. A control allocation algorithm is developed to cope with the actuator fault. Firstly, a nonlinear model of the quadrotor unmanned aerial vehicle (UAV) is presented. Then, the controller design and type of the actuator fault are explained. The control allocation algorithm is used to optimize the trajectory tracking performance of the quadrotor in the presence of an actuator fault. A control allocation algorithm is an effective approach to implementing fault-tolerant control. When actuator faults are identified, they can be modeled as changes in the B matrix of constraints. Various simulations have been made for situations with and without actuator failure. In normal conditions, the quadrotor can accurately track altitude, roll, pitch and yaw references. In faulty conditions, the quadrotor can follow the references with a small error. Simulations prove the effectiveness of the control allocation algorithm, which stabilizes the quadrotor in case of an actuator fault. Overall, this paper presents a novel fault-tolerant controller design for quadrotor aircraft that effectively addresses actuator faults using a super-twisting sliding mode controller and control allocation algorithm.

## 1. Introduction

The first quadrotors began to be produced in the 1920s. Produced from the 1920s to the 1970s, quadrotors were large-scale aircraft controlled by a pilot. With the development of technology, quadrotor unmanned aerial vehicles have been produced since the late 1990s. While the first quadrotors used by the pilot were produced for the purpose of carrying people and cargo, quadrotor UAVs can be used for many different purposes [1-3]. In the last few years, the importance of quadrotor UAVs has increased due to their extensive range of applications, such as monitoring, firefighting, crop spraying, mapping, transporting goods, aerial photography and military operations. In general, quadrotor UAVs can be used in various civil operations and defense industry applications [4-7]. The wide range of use, low cost and simple mechanical structure of quadrotor unmanned aerial vehicles have led many researchers to be interested in this issue [8]. Quadrotor unmanned aerial vehicles' vertical take-off and landing (VTOL) ability, hovering in the air, ability to rotate around their own axis, take-off and landing on uneven grounds make them more advantageous than winged unmanned aerial vehicles [9-12].

In practical applications, quadrotors are generally required to track a prescribed path accurately and rapidly. However, it is a challenging task because of parameter uncertainties, actuator faults and external disturbances [13]. Therefore, fault-tolerant controller design gains great importance for quadrotor unmanned aerial vehicles.

In the early studies, researchers mainly focused on PID controller design and linearized quadrotor dynamics, neglected some aerodynamic coefficients and ignored faults. Salih et al. designed a PID controller for a quadrotor UAV and neglected the inertia matrix of the aircraft and the gyroscopic effect caused by the propellers' rotation [14]. Ma et al. proposed a fuzzy PID controller for a quadrotor UAV, linearized the nonlinear system and neglected drag coefficients, aerodynamic, Coriolis and gyroscopic effects [15]. Shang et al. designed a fractional-order PID controller for a quadrotor UAV and ignored some small coefficients in the transfer function of the system [16]. Khatoon et al. compared PID and LQR controllers for a quadrotor UAV and linearized system about a certain operating point for using an LQR controller. It is concluded that the LQR controller is more robust and generates low steady-state error [17]. Huang et al. developed a backstepping controller, ignored some small coefficients related to inertia moments during the controller design process and compared results with a classical PID controller. It is observed that the backstepping controller has a lower settling time and faster time response than the PID controller [18]. Yu et al. developed an integral sliding mode controller for a quadrotor UAV and replaced the signum function with the saturation function to minimize the chattering effect in the controller's equations. Yu et al. also compared simulation results with the traditional PID controller and observed that the sliding mode controller converged rapidly to the attitude reference and showed better performance [19].

In the last few years, researchers have focused on fault-tolerant controller design for unmanned aerial vehicles. Fault-tolerant controllers maintain a stable performance regardless of disturbances [20], [21]. Ermeydan et al. designed a fault-tolerant PID controller for a quadrotor UAV. In the first and third rotors, a 20% loss of control effectiveness is simulated. It is observed that the PID controller can control the quadrotor stably up to 20% failure level in two rotors [22]. Sadeghzadeh et al. developed an active fault-tolerant PID controller for a quadrotor UAV, simulated a faulty scenario with up to 22% of power loss in power of all rotors and developed a gain scheduling strategy for faulty conditions [23]. Jun et al. proposed a fault-tolerant PID controller design for a quadrotor. It is observed that quadrotor flight is stable with a 2.2% height reduction when a 50% loss of effectiveness at one rotor [24]. Zhang et al. proposed linear quadratic regulator (LQR) and model predictive controllers (MPC) for a quadrotor UAV against rotor failures, considered 30% and 50% of loss effectiveness in one rotor, respectively and compared the results [25]. Nguyen et al. designed a fault-tolerant backstepping controller and performed simulations at 40% loss of control effectiveness at one rotor [26]. Some researchers have focused on developing a fault-tolerant sliding mode controller for the quadrotor UAV. Wang et al. developed an incremental sliding mode controller for a quadrotor UAV and simulated a 25% fault case for one rotor [27]. Yang et al. designed a sliding mode fault-tolerant controller for a quadrotor UAV with varying load and actuator faults. They proposed a method, a 40% power loss in one rotor at the 12th second [28]. Nguyen et al. developed a fault-tolerant sliding mode controller for a quadrotor UAV. They implemented a 40% loss of control effectiveness into the actuator [29]. In practical applications of sliding mode controllers, researchers may face the undesirable phenomenon of oscillations having finite frequency and amplitude, which is called chattering. Chattering leads to low control accuracy, high heat losses in power circuits and high wear of moving mechanical parts. It can arise from the fast dynamics which were neglected in the ideal model [30]. When a fault occurs in the system, the control allocation mechanism redistributes (re-allocates) the control signals produced by the controller among available actuators so that the controller has the same effects on the system output or at least produces the same steady-state responses [31].

In this research, a fault-tolerant super-twisting sliding mode controller with control allocation is proposed for a quadrotor UAV. 60% loss of control effectiveness at one rotor is simulated. The control allocation algorithm is used to stabilize the controller in case of an actuator fault. MATLAB simulations are performed for the cases with and without rotor failure, and the difference is observed. The major contributions of this research are outlined as follows:

1) Fault-tolerant super-twisting sliding mode controller with a control allocation algorithm is developed to control the quadrotor in case of actuator fault. The sliding-mode controllers are more robust than the other controller methods described above, but they suffer from a chattering problem. Unlike classical sliding mode controllers, our proposed controller does not have a chattering problem.

2) Many researchers have studied actuator failure at rates ranging from 20% to 50%. Our study focused on an actuator failure rate of 60%, higher than many other studies. Thanks to the control allocation algorithm, the quadrotor is controlled in a stable manner despite the high rate of actuator failure.

The remainder of this paper is structured as follows. In Section II, the dynamics of the quadrotor unmanned aerial vehicle are explained. In Section III, the fault-tolerant super-twisting sliding mode control structure is explained. The MATLAB simulations of the proposed fault-tolerant sliding mode controller with control allocation are presented in Section IV. Lastly, the conclusions of this research are stressed in Section V.

## 2. Quadrotor Model

Quadrotor is a four-rotor unmanned aerial vehicle with vertical landing and take-off features. The power unit of the quadrotor consists of two opposing rotor pairs. The (1,3) rotor pair and (2,4) rotor pair rotate in opposite directions to each other. There are four different movements in the quadrotor UAV. Simultaneously increasing or reducing the speed of all four rotors of the quadrotor leads to vertical movement [32]. Changing the speed of the (1,3) rotor pair of the quadrotor creates roll motion. Raising or reducing the speed of the (2,4) rotor pair of the quadrotor creates the pitch motion. The difference in torque between the (1,3) and (2,4) rotor pairs produce the yaw motion. A schematic representation of the four-rotor unmanned aerial vehicle with 6 Degrees of Freedom (DOF) is given in Fig. 1 [33]. The rotation direction of the rotors, the torques produced by the rotors, the Earth frame, the body frame, roll, pitch and yaw movements are represented in Fig. 1.

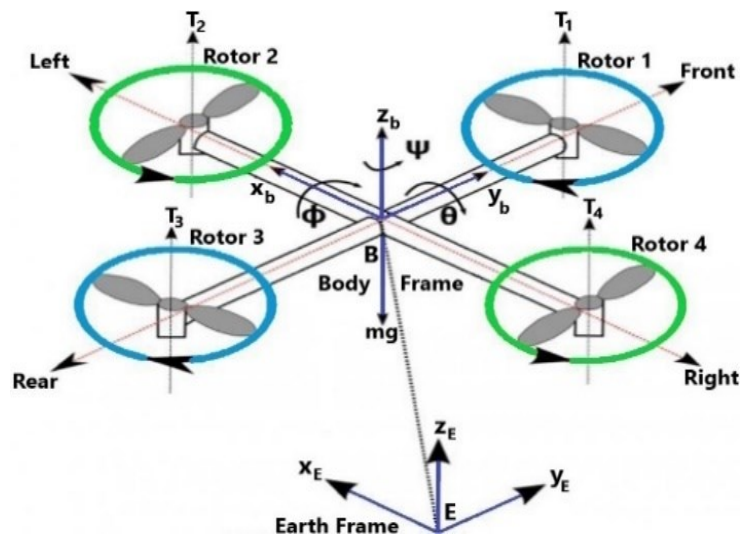

**Fig. 1.** Diagram of the quadrotor UAV

The quadrotor unmanned aerial vehicle with 6 DOF $[\varphi, \theta, \psi, x, y, z]$ consists of translational and rotational dynamics. Translational dynamics represent $x, y, z$ axes, whereas rotational dynamics $(\varphi, \theta, \psi)$ present respectively roll, pitch and yaw. The Earth frame is described $E = [X_E, Y_E, Z_E]$ and the body frame is given as $B = [X_B, Y_B, Z_B]$ whose origin B defines the center of gravity in the quadrotor UAV [32]. The rotor is defined as a propulsion system, and it provides the thrust with a propeller [34]. In Fig. 2, the connection between the rotor and the propeller appears more clearly. Also, in Fig. 2, roll, pitch and yaw movements are shown on the axes [35].

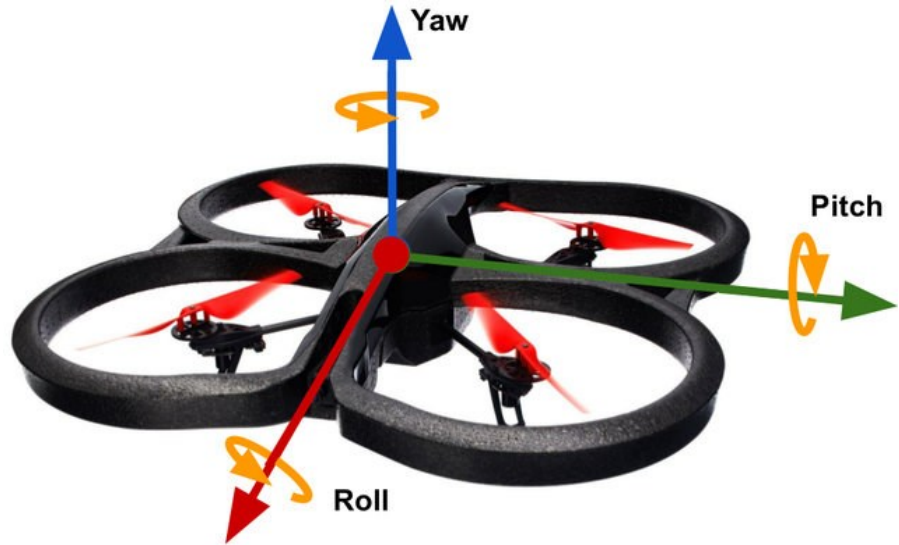

**Fig. 2.** Rotor-propeller connection and rotational movements of the quadrotor UAV

Rotation matrix R is defined in (1), and it is used with the aim of transformation between Earth frame E and body frame B.

$$R = \begin{bmatrix} c\theta c\psi & s\varphi s\theta c\psi - c\varphi s\psi & c\varphi s\theta c\psi + s\varphi s\psi \\ c\theta s\psi & s\varphi s\theta s\psi + c\varphi c\psi & c\varphi s\theta s\psi - s\varphi c\psi \\ -s\theta & s\varphi c\theta & c\varphi c\theta \end{bmatrix} \quad (1)$$

where

$$c = \cos, \quad s = \sin \quad (2)$$

$I$ matrix represents the Inertia matrix and includes diagonal elements with their product equal to 0, having $I_x$, $I_y$, $I_z$ moments of inertia. The matrix of Inertia is given in (3).

$$I = \begin{bmatrix} I_x & 0 & 0 \\ 0 & I_y & 0 \\ 0 & 0 & I_z \end{bmatrix} \quad (3)$$

The forces produced by rotors and moments have been explained by (4) and (5) as:

$$F_i = K_f \Omega_i^2 \quad (4)$$

$$M_i = K_m \Omega_i^2 \quad (5)$$

where $K_f$ and $K_m$ are constant coefficients, $\Omega_i^2$ defines the i$^{th}$ angular velocity. The actuator dynamics are represented in (6). $U_1$, $U_2$, $U_3$, $U_4$ are control inputs. $F$ describes the total thrust produced by the propellers. $T_\varphi$, $T_\theta$, $T_\psi$, are torques along the respective axes. $l$ presents the arm length of the quadrotor.

$$u = \begin{bmatrix} U_1 \\ U_2 \\ U_3 \\ U_4 \end{bmatrix} = \begin{bmatrix} F \\ T_\varphi \\ T_\theta \\ T_\psi \end{bmatrix} = \begin{bmatrix} K_f & K_f & K_f & K_f \\ 0 & -lK_f & 0 & lK_f \\ lK_f & 0 & -lK_f & 0 \\ -K_m & K_m & -K_m & K_m \end{bmatrix} \begin{bmatrix} \Omega_1^2 \\ \Omega_2^2 \\ \Omega_3^2 \\ \Omega_4^2 \end{bmatrix} \quad (6)$$

The quadrotor's dynamical model is defined as

$$\ddot{\varphi} = \dot{\psi}\dot{\theta}(\frac{I_y - I_z}{I_x}) + \dot{\theta}\Omega_R \frac{J_R}{I_x} + \frac{l}{I_x} U_2 \quad (7)$$

$$\ddot{\theta} = \dot{\psi}\dot{\theta}(\frac{I_z - I_x}{I_y}) + \dot{\varphi}\Omega_R \frac{J_R}{I_x} + \frac{l}{I_y} U_3 \quad (8)$$

$$\ddot{\psi} = \dot{\varphi}\dot{\theta}(\frac{I_x - I_y}{I_z}) + \frac{l}{I_z} U_4 \quad (9)$$

$$\ddot{X} = (\cos\varphi \sin\theta \cos\psi + \sin\psi \sin\varphi) \frac{U_1}{m} \quad (10)$$

$$\ddot{Y} = (\cos\varphi \sin\theta \sin\psi - \cos\psi \sin\varphi) \frac{U1}{m} \quad (11)$$

$$\ddot{Z} = (\cos\theta \cos\varphi) \frac{U_1}{m} - g \quad (12)$$

where $\Omega_R$ and $J_R$ are angular velocity and inertia moment of the propellers, $m$ describes the quadrotor's mass, and $g$ presents gravitational acceleration.

In this research, MATLAB simulations are performed using the physical properties of the Parrot AR Drone 2.0 quadrotor UAV. It has two interchangeable hulls. One is designed for indoors, and the other for outdoors. The hull designed for indoor use is heavier but more resistant to collisions. In this study, the indoor hull is used. Fig. 3 shows the Parrot AR Drone 2.0 with the indoor hull [36]. Table 1 represents the parameters of the Parrot AR Drone 2.0 with the indoor hull [37, 38].

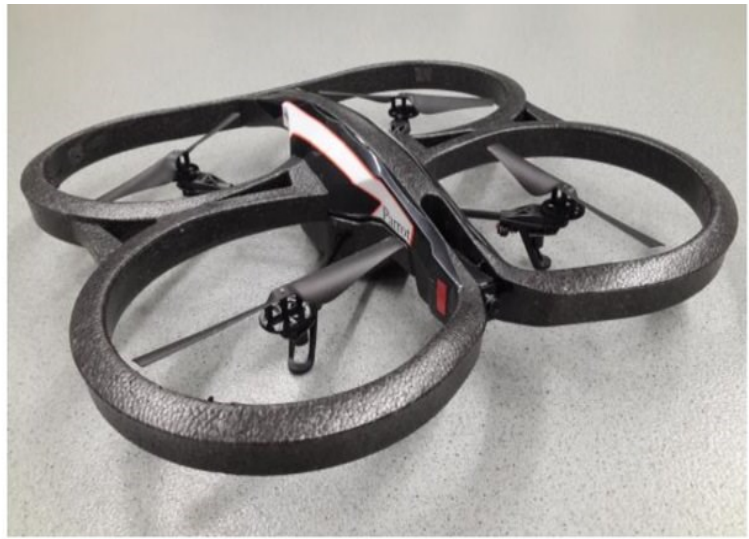

**Fig. 3.** Parrot AR Drone 2.0 with indoor hull

Table 1. Quadrotor parameters

| Symbol | Quantity | Value |
|---|---|---|
| $m$ | Mass of the quadrotor | 0.429 kg |
| $l$ | Arm length of the quadrotor | 0.1785 m |
| $J_R$ | The inertia of each rotor | 2.03 x $10^{-5}$ kg·m² |
| $I_x$ | Inertia for $x$-axes | 2.24 x $10^{-3}$ kg·m² |
| $I_y$ | Inertia for $y$-axes | 2.98 x $10^{-3}$ kg·m² |
| $I_z$ | Inertia for $z$-axes | 4.80 x $10^{-3}$ kg·m² |
| $K_f$ | Thrust factor | 8.05 × $10^{-6}$ N/(rad/s)² |
| $K_m$ | Torque factor | 2.42 × $10^{-7}$ N·m/(rad/s)² |
| $\Omega_{max}$ | Maximum speed | 1047.2 rad/s |

## 3. Controller Design

The fault-tolerant super twisting sliding mode control is proposed to track the given altitude and attitude references in case of 60% actuator failure at one rotor. The control allocation algorithm is activated when one of the four rotors fails and enables the quadrotor to follow the references stably.

### 3.1. Super Twisting Sliding Mode Controller

The sliding mode controller design is more robust than other controller methods in the literature. Two stages are very important in the sliding mode controller approach.

1) The selection of the sliding surface

2) State variables must remain on the sliding surface

The sliding phase and the reaching phase of the sliding mode controller are represented in Fig. 4. and the sliding surface is defined in (13) [39].

$$s = \dot{e} + \lambda e \quad (13)$$

where $e$ is the error and $\lambda$ represents the tuning parameter.

The sliding variables used in the sliding mode controller are given in (14).

$$s^T = [\varphi\ \theta\ \psi\ z] \quad (14)$$

The sliding variables must reach the sliding surface in finite time and slide on the surface to reach 0 asymptotically with the aim of controlling the quadrotor to the reference trajectory [40].

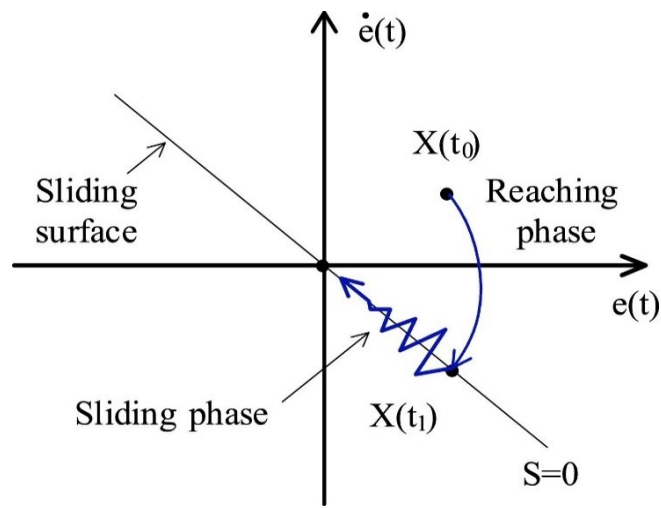

Fig. 4. Sliding phase and reaching phase

The equivalent control inputs for the sliding mode controller can be summarized as follows.

$$U_{1eq} = \frac{m}{\cos\varphi \cos\theta}(g + \ddot{z}_d - \lambda(\dot{z}_d - \dot{z})) \tag{15}$$

$$U_{2eq} = \frac{I_x}{l}\left(-\dot{\theta}\dot{\psi}\left(\frac{I_y - I_z}{I_x}\right) + \ddot{\varphi} - \lambda(\dot{\varphi}_d - \dot{\varphi})\right) \tag{16}$$

$$U_{3eq} = \frac{I_y}{l}\left(-\dot{\varphi}\dot{\psi}\left(\frac{I_z - I_x}{I_y}\right) + \ddot{\theta} - \lambda(\dot{\theta}_d - \dot{\theta})\right) \tag{17}$$

$$U_{4eq} = \frac{I_z}{l}\left(-\dot{\varphi}\dot{\theta}\left(\frac{I_x - I_y}{I_z}\right) + \ddot{\psi}_d - \lambda(\dot{\psi}_d - \dot{\psi})\right) \tag{18}$$

A discontinuous term is added to the equivalent control for the purpose of satisfying the sliding reachability condition by substituting

$$\dot{s} = -\,sign(s) \tag{19}$$

The equivalent control equation can be written as follows.

$$U_1 = U_{1eq} + sign(s_z) \tag{20}$$

$$U_2 = U_{2eq} + sign(s_\varphi) \tag{21}$$

$$U_3 = U_{3eq} + sign(s_\theta) \tag{22}$$

$$U_4 = U_{4eq} + sign(s_\psi) \tag{23}$$

The sign function being discontinuous will cause a high-frequency oscillation called the chattering effect [41]. In order to cope with the chattering effect sign function is replaced with an alternative continuous function (24).

$$fun(s) = [(1+s)^n - (1-s)^n]/[(1+s)^n + (1-s)^n] \tag{24}$$

where $n$ is an even integer greater than 0, signum function and alternative continuous function graphs are shown in Fig. 5 through Fig. 6. Fig. 5 represents the alternative continuous function when $n = 2$.

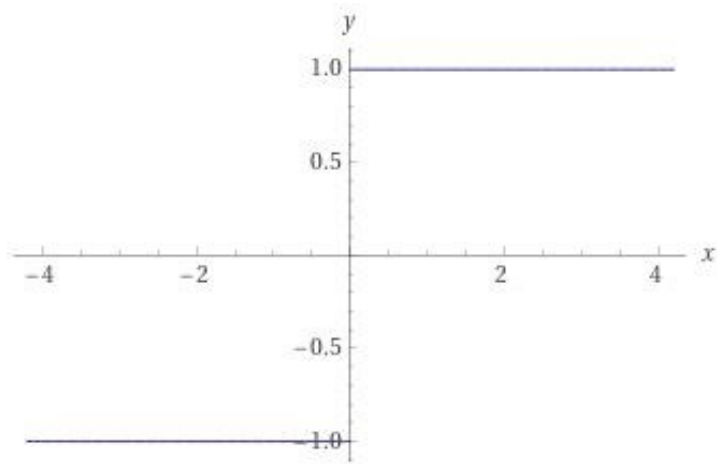

**Fig. 5.** The signum function

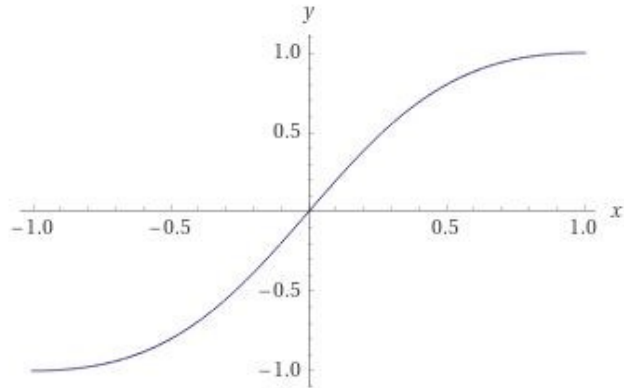

**Fig. 6.** A continuous function with n = 2

The use of the continuous function reduces chattering; however, it can lead to decreasing the robustness of the controller. A higher-order linear switching function can be replaced instead of a nonlinear switching function with the purpose of reducing the chattering effect. Super-twisting sliding mode controller presents such a switching function that can balance with the robustness of the controller and high-frequency chattering.

In order for the sliding variables to reach the sliding surface in a finite time and without chattering, the following higher-order switching function is used.

$$\dot{S} = 1.5k^{1/2}s^{1/2}\,sign(s) + \int 1.1\,ksign(s) \tag{25}$$

The control inputs for the super-twisting sliding mode controller are calculated in Equations (26)-(29) where $k_1, \ldots, k_4$ are positive constants.

$$U_1 = U_{1eq} - 1.5k_1^{1/2}s_z^{1/2}\,sign(s_z) - \int 1.1\,k_1 sign(s_z) \tag{26}$$

$$U_2 = U_{2eq} - 1.5k_2^{1/2}s_\varphi^{1/2}\,sign(s_\varphi) - \int 1.1\,k_2\,sign(s_\varphi) \tag{27}$$

$$U_3 = U_{3eq} - 1.5k_3^{1/2}s_\theta^{1/2}\,sign(s_\theta) - \int 1.1\,k_3\,sign(s_\theta) \tag{28}$$

$$U_4 = U_{4eq} - 1.5k_4^{1/2}s_\psi^{1/2}\,sign(s_\psi) - \int 1.1\,k_4\,sign(s_\psi) \tag{29}$$

### 3.1. Control Allocation Algorithm

When one of the rotors loses its effectiveness by 60%, the controller should counter the loss of effectiveness and continue the reference track safely. The loss of actuator effectiveness is calculated by comparing the desired angular velocity of the rotor ($w_d$) with the actual angular velocity of the rotor ($w$). The loss of effectiveness ($LE$) is calculated as in (30).

$$LE = (w_d - w)/w_d \tag{30}$$

The LE value is always between 0 and 1. When the $LE$ value is 0, there is no power loss in the rotor, and when it is 1, there is a 100% power loss in the rotor. When the $LE$ value rises above 0, the control allocation algorithm is activated, and the power loss in the rotor is compensated. When one of the propellers of the quadrotor fails, the quadrotor loses its balance. When the faults are determined, the angular velocity of the rotor opposite to the faulty rotor is calculated such that the torque balance is maintained. The control effectiveness of the quadrotor UAV is defined in (31).

$$k = 1 - LE \qquad (31)$$

The control effectiveness vector is described in (32).

$$K = \begin{bmatrix} k_1 \\ k_2 \\ k_3 \\ k_4 \end{bmatrix} \qquad (32)$$

where $k_i \in [0\ 1]$ and $k_i = 1$ means that the actuator is completely functional and $k_i = 0$ describes the case where the actuator does not work at all. After the fault is detected, the control output is represented as

$$\begin{bmatrix} F \\ T_\varphi \\ T_\theta \\ T_\psi \end{bmatrix} = \begin{bmatrix} K_f & K_f & K_f & K_f \\ 0 & -lK_f & 0 & lK_f \\ lK_f & 0 & -lK_f & 0 \\ -K_m & K_m & -K_m & K_m \end{bmatrix} \begin{bmatrix} \Omega_1^2 \\ \Omega_2^2 \\ \Omega_3^2 \\ \Omega_4^2 \end{bmatrix} \qquad (33)$$

The fault-tolerant super twisting sliding mode controller diagram is represented in Fig. 7. Sliding mode control is a variable structure control method. The state feedback control law is not a continuous function of time. It can switch from one continuous structure to another continuous structure based on the current position in the state space [42], [43]. The sliding variable must be used to define the sliding surface. The reference trajectory is the trajectory given for the quadrotor to follow. The quadrotor tries to follow this trajectory using the sliding mode controller. When an actuator error occurs, and power loss occurs in one of the rotors, the control allocation mechanism is activated. It readjusts the power distribution between the rotors and keeps the quadrotor flying stably.

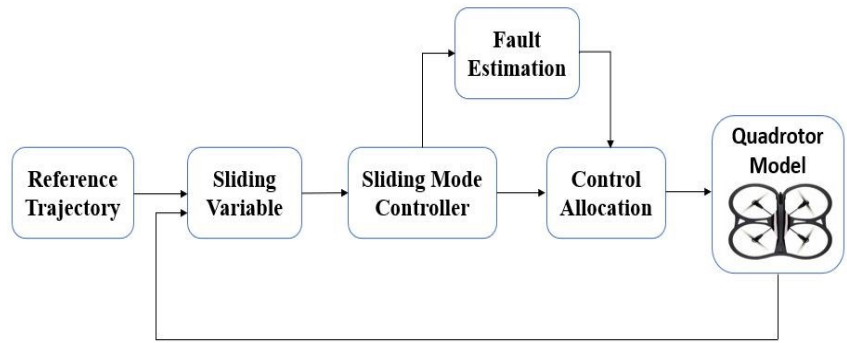

**Fig. 7.** Block scheme of the proposed controller

## 4. Result and Discussions

In this part, the efficiency of the proposed fault-tolerant super-twisting sliding mode controller is verified by MATLAB simulations. The physical parameters of the quadrotor defined in the second section are used in the MATLAB simulations. An actuator fault due to 60% power loss in the fourth rotor is introduced at the beginning of the simulation. Simulations were made separately for cases with and without actuator failure. In this way, it has been observed more clearly how tolerant the proposed controller is against actuator errors. Fig. 8-11 shows the roll, pitch, yaw and altitude tracking responses of the controller without actuator failure.

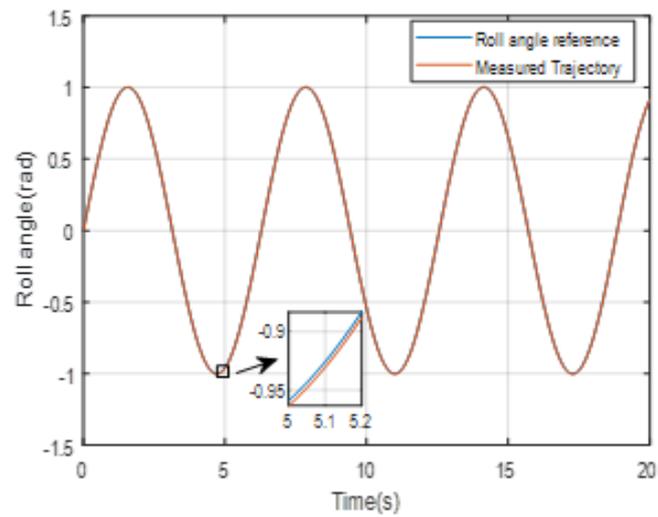

**Fig. 8.** Roll angle reference tracking simulation of the controller without rotor failure

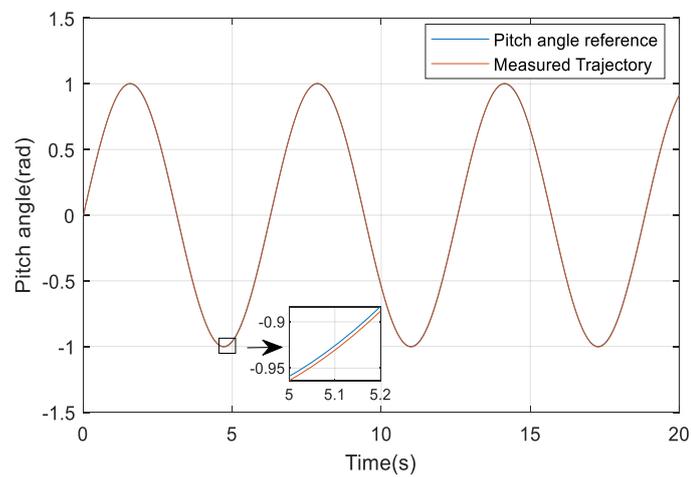

**Fig. 9.** Pitch angle reference tracking simulation of the controller without rotor failure

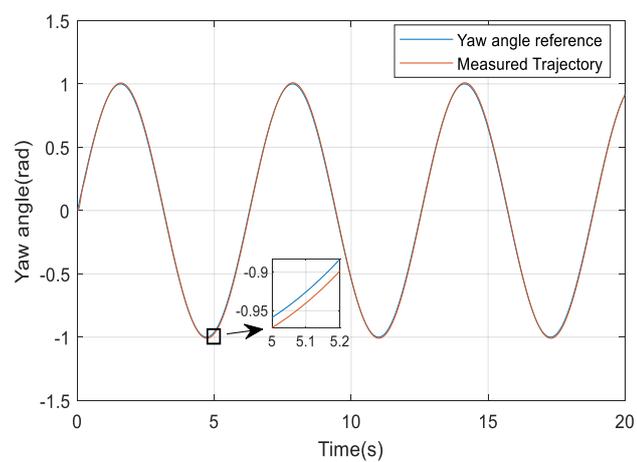

**Fig. 10.** Yaw angle reference tracking simulation of the controller without rotor failure

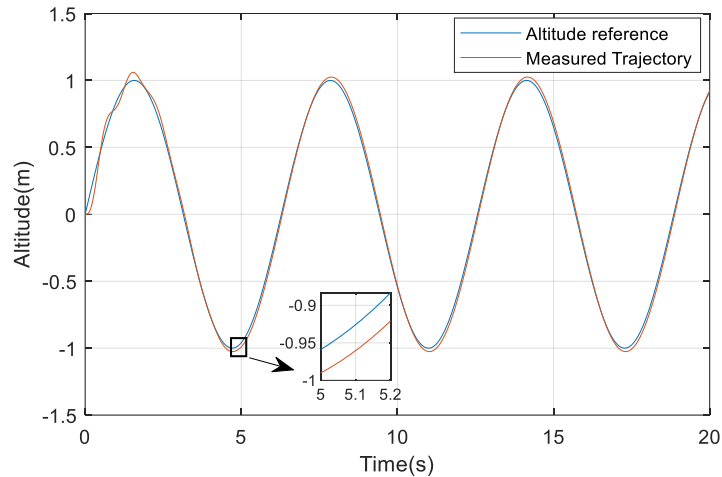

**Fig. 11.** Altitude reference tracking simulation of the controller without rotor failure

It is clear from the simulations in Fig. 8-11 that the quadrotor can follow the given sinusoidal references with almost 0 error. The references for roll, pitch, and yaw angles are in radians, and the reference for altitude is in meters. The red line in the graphs represents the given sinusoidal reference, while the blue line shows the trajectory followed by the quadrotor. The part between 5 and 5.2 seconds in the graphs has been enlarged to provide a clearer observation of the reference (red line) and the trajectory followed by the quadrotor (blue line).

Table 2 gives the time response (rise time, overshoot and settling time) values of the controller without rotor failure. Rise time is defined as the time required for a signal to change from 10% to 90% of a specified value. Overshoot is the occurrence of a signal exceeding its target. Settling time is the time required by the response to reach and steady within a specified range of 2 % of its final value. When the data in Table 2 is carefully examined, it is seen that the references of roll, pitch and yaw angles are followed with a slight overshoot of around 0.5%. The altitude reference is followed by an overshoot of 2.2%. The controller followed the roll, pitch and yaw references very closely from the start of the simulation and never left the ± 2% range. Therefore, the settling time is 0 seconds for these references. The altitude reference shows a little more overshoot and settles in the ±2% range in 0.25 seconds. In the absence of rotor error, the controller can follow the given references almost without error.

**Table 2.** Time response of the controller without rotor failure

| Reference | Rise time (s) | Overshoot (%) | Settling time (s) |
| --- | --- | --- | --- |
| Roll angle | 1.828 | 0.503 | 0 |
| Pitch angle | 1.828 | 0.503 | 0 |
| Yaw angle | 1.829 | 0.505 | 0 |
| Altitude | 1.818 | 2.2 | 0.25 |

Fig. 12-15 show simulations indicating the reference tracking results when the fourth rotor of the quadrotor experiences 60% power loss. In case of a rotor failure in the quadrotor, the control allocation algorithm activates, ensuring that the quadrotor stably follows the given references. The control allocation algorithm can compensate for the failure of the fourth rotor with the help of the other 3 rotors.

In case of 60% rotor failure, the quadrotor can still follow the given references. However, a small overshoot is observed at the peaks of the sinusoidal references. This is due to power loss in one rotor of the quadrotor. As explained in Section II, the torque difference between the rotor pairs (1,3) and (2,4) produces the yaw motion. Due to the power loss in the fourth rotor, the torque difference between the rotor pairs increases, and a higher overshoot occurs during the yaw angle reference tracking. When evaluated in general, it is clear that the control allocation algorithm works successfully and achieves stable reference tracking by compensating for rotor failure.

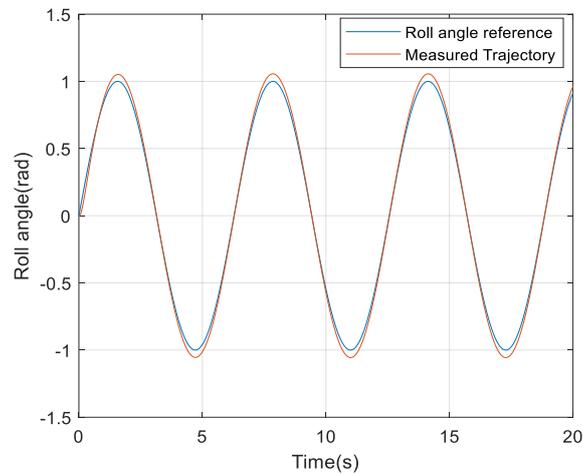

**Fig. 12.** Roll angle reference tracking simulation of the controller with rotor failure

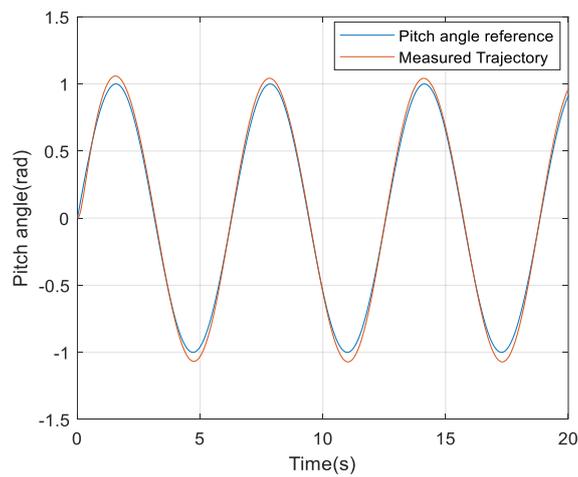

**Fig. 13.** Pitch angle reference tracking simulation of the controller with rotor failure

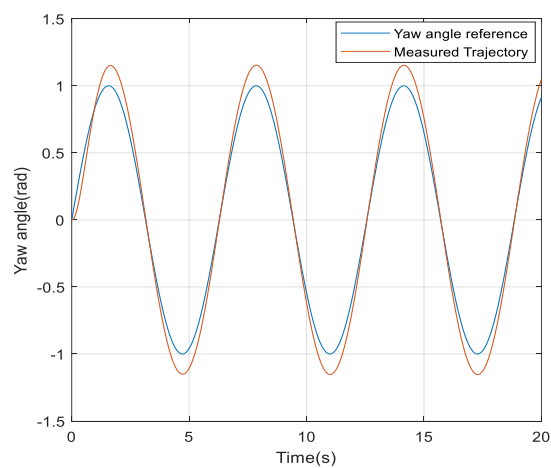

**Fig. 14.** Yaw angle reference tracking simulation of the controller with rotor failure

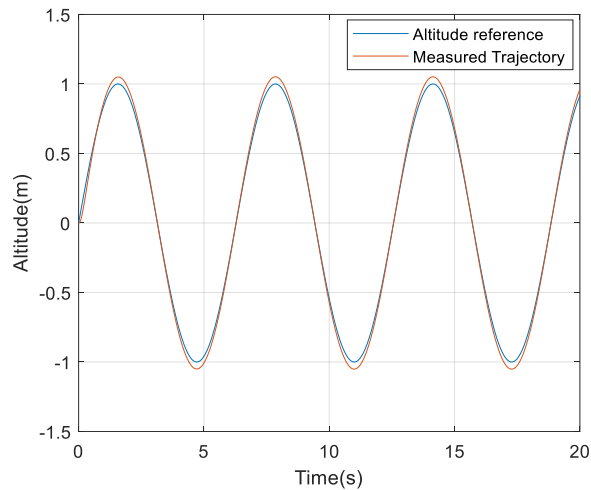

**Fig. 15.** Altitude reference tracking simulation of the controller with rotor failure.

Table 3 gives the time response (rise time, overshoot and settling time) values of the controller with rotor failure. When the time response data in Table 3 is examined, it is seen that the rotor error increases the overshoot. However, it is not high enough to prevent the quadrotor from flying stably. The settling time is slightly longer than when there is no rotor error. However, the reference value is captured at around 0.6 seconds for roll, pitch and altitude references. The yaw reference can be captured in about 1 second. The system has a short settling time despite rotor failure.

**Table 3.** Time response of the controller with rotor failure

| Reference | Rise time (s) | Overshoot (%) | Settling time (s) |
|---|---|---|---|
| Roll angle | 1.829 | 5.70 | 0.65 |
| Pitch angle | 1.807 | 5.72 | 0.60 |
| Yaw angle | 1.830 | 15.2 | 1.08 |
| Altitude | 1.829 | 4.80 | 0.64 |

## 5. Conclusion

This research proposes a fault-tolerant super-twisting sliding mode controller for a quadrotor. Firstly, the development of the quadrotor's dynamic model is explained. Then, the working principle of the proposed controller is explained. To eliminate the chattering effect in the sliding mode controller, the signum function has been replaced with a continuous function. Then, it is explained how the fault-tolerant controller is implemented by using the control allocation algorithm. Thanks to the control allocation algorithm, 60% power loss in a single rotor can be tolerated. In this way, the quadrotor can follow the given references in a fault-tolerant manner. In the simulations section, different simulations have been made for cases with and without rotor failure. In this way, the efficiency of the fault-tolerant sliding mode controller is better observed. Simulations were made for reference tracking, and the results were compared. It has been observed that in the absence of rotor failure, the quadrotor follows the sinusoidal references almost flawlessly. In the case of 60% rotor failure in a single rotor, it has been observed that the quadrotor follows the references stably thanks to the control allocation algorithm. In addition, rise time, overshoot and settling time data were obtained for cases with or without rotor failure. When the data obtained are examined, it is seen that the quadrotor successfully follows all the references, almost without overshoot, in the absence of rotor failure. When power loss occurs due to rotor error, the quadrotor can follow the references by showing a small overshoot. As a result, it has been observed that the fault-tolerant super-twisting sliding mode controller with control allocation works successfully. In future studies, a control allocation algorithm will be developed that will enable the quadrotor aircraft to fly stably in case of a failure of the two rotors.

**Author Contribution:** All authors contributed equally to the main contributor to this paper. All authors read and approved the final paper.

**Funding:** This research received no external funding.

**Acknowledgment:** We would like to express our sincere gratitude to our supervisor, Professor Cosku Kasnakoglu, for his valuable guidance and support throughout the research process. His expertise and insights were invaluable in shaping our research and helping us to overcome challenges.

**Conflicts of Interest:** The authors declare no conflict of interest.